\documentclass[aps,floatfix,prl,superscriptaddress,reprint,showpacs,10pt,preprintnumbers,longbibliography]{revtex4-1}
\usepackage[utf8]{inputenc}
\usepackage[pdftex]{graphicx}
\usepackage{float}
\usepackage{amsmath}
\usepackage{amssymb}
\usepackage{mathtools}
\usepackage{siunitx}
\usepackage{amsfonts}
\usepackage{dsfont}
\usepackage{array}
\usepackage{bm}
\usepackage{mathrsfs}
\usepackage{pifont}
\usepackage{multirow}
\usepackage{upgreek}
\usepackage{longtable,booktabs}
\usepackage[dvipsnames]{xcolor}
\usepackage[pdftex,
  pdftitle={Quark and gluon momentum fractions in the pion from Nf=2+1+1 lattice QCD},
  pdfauthor={C. Alexandrou, S. Bacchio, G. Bergner, J. Finkenrath,
    A. Gasbarro, K. Hadjiyiannakou, K. Jansen, B. kostryewa,
    K. Ottnad, M. Petschlies, F. Pittler, F. Steffens, C. Urbach, U. Wenger},
  bookmarks,
  colorlinks,
  linkcolor=myblue,
  citecolor=mymagenta,
  menucolor=black,
  urlcolor=myblue,
  plainpages=false,
  pdfpagelabels,
  hypertexnames=false]{hyperref}
\usepackage{verbatim}
\usepackage{slashed}
\usepackage[capitalise]{cleveref}
\usepackage{ucs}
\usepackage{subfigure}
\usepackage{csquotes}

\newcommand{\gev}{\,\mathrm{GeV}}

\newcommand{\gsim}{{\;\raise0.3ex\hbox{$>$\kern-0.75em\raise-1.1ex\hbox{$\sim$}}\;}}

\newcommand{\avgx}{\langle x\rangle}
\newcommand{\pvec}{\mathbf{p}}
\newcommand{\seeApp}{the appendix}

\definecolor{mymagenta}{RGB}{200, 0, 100}
\definecolor{myblue}{RGB}{45, 48, 146}

\graphicspath{{plots/}}

\begin{document}
\title{Quark and gluon momentum fractions in the pion from $N_f=2+1+1$
  lattice QCD}

\author{Constantia Alexandrou}
\affiliation{Department of Physics, University of Cyprus, Nicosia, Cyprus}
\affiliation{Computation-based Science and Technology Research Center, The Cyprus Institute, Nicosia, Cyprus}

\author{Simone Bacchio}
\affiliation{Computation-based Science and Technology Research Center, The Cyprus Institute, Nicosia, Cyprus}

\author{Georg Bergner}
\affiliation{Institute of Theoretical Physics, Friedrich Schiller
  University Jena, Germany}

\author{Jacob Finkenrath}
\affiliation{Computation-based Science and Technology Research Center, The Cyprus Institute, Nicosia, Cyprus}

\author{Andrew Gasbarro}
\affiliation{Albert Einstein Center, Institute for Theoretical Physics, University of Bern, Switzerland}

\author{Kyriakos Hadjiyiannakou}
\affiliation{Department of Physics, University of Cyprus, Nicosia, Cyprus}
\affiliation{Computation-based Science and Technology Research Center, The Cyprus Institute, Nicosia, Cyprus}

\author{Karl Jansen}
\affiliation{NIC, DESY Zeuthen, Germany}

\author{Bartosz Kostrzewa}
\affiliation{High Performance Computing and Analytics Lab, Rheinische Friedrich-Wilhelms-Universität Bonn,
Germany}

\author{Konstantin Ottnad}
\affiliation{PRISMA$^+$ Cluster of Excellence and Institut für
Kernphysik, Johannes Gutenberg-Universität Mainz, Germany}

\author{Marcus Petschlies}
\affiliation{Helmholtz-Institut für Strahlen- und Kernphysik, University of Bonn, Germany}
\affiliation{Bethe Center for Theoretical Physics, University of Bonn, Germany}

\author{Ferenc Pittler}
\affiliation{Computation-based Science and Technology Research Center, The Cyprus Institute, Nicosia, Cyprus}

\author{Fernanda Steffens}
\affiliation{Helmholtz-Institut für Strahlen- und Kernphysik, University of Bonn, Germany}
\affiliation{Bethe Center for Theoretical Physics, University of Bonn, Germany}

\author{Carsten Urbach}
\affiliation{Helmholtz-Institut für Strahlen- und Kernphysik, University of Bonn, Germany}
\affiliation{Bethe Center for Theoretical Physics, University of Bonn, Germany}

\author{Urs Wenger}
\affiliation{Albert Einstein Center, Institute for Theoretical Physics, University of Bern, Switzerland}
\affiliation{Department of Theoretical Physics, CERN, Geneva, Switzerland}

\collaboration{Extended Twisted Mass Collaboration}
\date{\today}
\begin{abstract}
  We perform the first full decomposition of the pion momentum into
  its gluon and quark contributions. We employ an ensemble generated
  by the Extended Twisted Mass Collaboration with $N_f=2 + 1 +1$
  Wilson twisted mass clover fermions at maximal twist tuned to
  reproduce the physical pion mass. We present our results in the
  $\overline{\mathrm{MS}}$ scheme at $2\gev$. We find
  $\avgx_{u+d}=0.601(28)$, $\avgx_s=0.059(13)$, $\avgx_c=0.019(05)$,
  and $\avgx_g=0.52(11)$ for the separate contributions, 
  respectively, whose sum saturates the momentum sum rule.
\end{abstract}

\maketitle

\textit{Introduction.---}
Quantum Chromodynamics (QCD) manifests itself in the form of a
plethora of states -- so called hadrons, formed by quarks and gluons.
Pions are particularly interesting hadrons: they are
the lightest and simplest of the QCD bound states composed out of
quark and antiquark. At the same time pions are also pseudo-Goldstone
bosons, with the spontaneous breaking of chiral symmetry playing a
fundamental role in the emergence of their mass. Yet, in contrast to
the nucleon (proton and neutron), a first
principles computation of the pion structure, and in particular how
quarks and gluons contribute to its mass and momentum decomposition is
still lacking. The importance of this topic is well represented in the
Electron Ion Collider (EIC) yellow report~\cite{AbdulKhalek:2021gbh}:
eight main science questions concerning pions (and kaons) are prominently
put forward. Let us highlight two of these questions here:
\enquote{What are the quark and gluon energy
contributions to the  pion mass?}, and \enquote{Is the pion full or
  empty of gluons as viewed at large $Q^2$?}
The results presented in this letter on the momentum decomposition of
the pion using lattice QCD simulations address both
questions.

As mentioned before, there is a wealth of studies on the nucleon
momentum decomposition available in the literature using
phenomenological analyses of experimental
data~\cite{Ball:2017nwa,Dulat:2015mca,Alekhin:2017kpj,Moffat:2021dji},
and, more recently, from precise simulations of lattice QCD at the physical
point~\cite{Alexandrou:2017oeh,Alexandrou:2020sml}. The reason for the
pion being much less well investigated is that proton and neutron
structure is experimentally well accessible, while the pion is
significantly more challenging because there is no pion target available.
For that reason, only recently the first Monte Carlo global QCD
analysis for pion PDFs has been presented in
Ref.~\cite{Barry:2018ort}, which includes leading neutron
electroproduction (LNE) data from HERA and Drell-Yan data from CERN
and Fermilab. One of their interesting findings is that the
decomposition of the pion momentum, $\langle x \rangle_\pi$, into its
valence, $\langle x \rangle_v$, sea, $\langle x \rangle_s$, and gluon,
$\langle x \rangle_g$ components depends strongly on which data set is
included in the analysis. 
In particular, the inclusion of LNE data, which induces a model dependence
in the extraction of the pion PDFs, has a significant effect on the average
momentum carried by gluons and sea quarks in the pion. Precise lattice QCD
data for both quark and gluon momentum fractions has, thus, the potential
to add new model independent constraints on the extraction of pion PDFs from
experimental data. Finally, new data coming
from planned EIC experiments, as well as from COMPASS++/Amber~\cite{Denisov:2018unj}
will help to clarify the quark and gluon dynamics within the pion.

On the theory side one has to resort to
models~\cite{Ding:2019lwe,Freese:2021zne} or to nonperturbative
methods as provided by lattice QCD. Also from the lattice side,  there
is surprisingly little known for the pion. Most of the computations
available so
far~\cite{Martinelli:1987zd,Best:1997qp,Guagnelli:2004ga,Capitani:2005jp,Abdel-Rehim:2015owa,Oehm:2018jvm,Alexandrou:2020gxs,Alexandrou:2021mmi}
neglect potentially  important contributions, the so-called
quark disconnected contributions. While we were finalising the
present work a first computation including disconnected contributions
was put forward~\cite{Loffler:2021afv}.
Also for the gluon contributions there exists only a
single computation and only in the quenched
approximation~\cite{Meyer:2007tm}. Thus, systematics are certainly not
sufficiently controlled. More recently, there are
studies using
quasi-~\cite{Ji:2013dva,Chen:2018fwa,Lin:2020ssv,Gao:2020ito} and
pseudo-distributions~\cite{Radyushkin:2017cyf,Joo:2019bzr} as well as
so-called good lattice cross  
sections~\cite{Ma:2017pxb,Sufian:2019bol,Sufian:2020vzb} approaches 
to compute the $x$ dependence of the pion PDFs directly on the
lattice. These studies, however, are restricted to connected contributions
only.

In this letter we present the first calculation of the quark and gluon
momentum fractions in the pion based on lattice QCD simulations with $N_f=2+1+1$
dynamical quark flavours including all required contributions. The
computation is performed using one ensemble with physical values of
all four quark mass parameters. This allows us to check the momentum
sum rule, i.e.
whether all four quark and the gluon fractions sum up to
one. This result can pave the way towards a global QCD analysis
including experimental data as well as lattice QCD results of the
pion, which will help to sort out the discrepancy found
between different experimental data sets. 

\textit{Lattice Computation.---} Our computation is based on an
ensemble~\cite{Alexandrou:2018egz} 
generated by the Extended Twisted Mass Collaboration (ETMC) using
$N_f=2+1+1$ dynamical Wilson twisted mass clover fermions at maximal 
twist~\cite{Frezzotti:2000nk,Frezzotti:2003xj} and Iwasaki gauge
action~\cite{Iwasaki:1985we}. With this discretisation, lattice
artefacts are of $O(a^2)$ only~\cite{Frezzotti:2003ni}.
The lattice volume is $L^3\times T=64^3\times128$ and the lattice spacing
$a=0.08029(41)\ \mathrm{fm}$.
For strange and charm quarks we use a mixed action approach following
Ref.~\cite{Frezzotti:2004wz},
and all quark mass parameters are tuned to assume approximately
physical values \cite{Alexandrou:2018egz,Alexandrou:2021gqw}.
We give further details on quark mass tuning in the appendix.
For all estimates we used 745 well-separated gauge configurations.

The relevant elements of the traceless Euclidean energy-momentum
tensor (EMT)
for quark flavour $q$ with the symmetrised covariant derivative
$\stackrel{\leftrightarrow}{D}_\mu$ read
\begin{equation}
\bar{T}_{\mu\nu}^q\ =\ -\frac{(i)^{\kappa_{\mu\nu}}}{4}\,\bar q\,
  \left(
    \gamma_\mu\stackrel{\leftrightarrow}{D}_\nu + \gamma_\nu \stackrel{\leftrightarrow}{D}_\mu
    - \delta_{\mu\nu}\,\frac{1}{2}\,\gamma_\rho\,\stackrel{\leftrightarrow}{D}_\rho
  \right)q\,,
\end{equation}
with $\kappa_{\mu\nu} = \delta_{\mu,4}\,\delta_{\nu,4}$.
Analogously for the gluon 
\begin{equation}
\bar{T}^g_{\mu\nu} \ =\ (i)^{\kappa_{\mu\nu}}\,\left( 
  F_{\mu\rho}^{~}F_{\nu\rho} +
  F_{\nu\rho}^{~}F_{\mu\rho} - \delta_{\mu\nu}\,\frac{1}{2}\,F_{\rho\sigma}{~}F_{\rho\sigma}
  \right) \,.
\end{equation}
For $X=u, d, s, c, g$, one then obtains $\langle x\rangle^X$ from
\begin{equation}
  \label{eq:x}
  \langle\pi(\pvec)|\bar{T}^X_{\mu\nu}|\pi(\pvec)\rangle\ =\
  2\langle x\rangle^X\left(p_\mu p_\nu - \delta_{\mu\nu}\frac{p^2}{4}\right)
\end{equation}
with on-shell momentum $p = \left( E_\pi=\sqrt{m_\pi^2 + \pvec^2},\,\pvec \right)$.
We extract these matrix elements from ratios of Euclidean three- and
two-point functions 
\begin{equation}
  \label{eq:Rdef}
  R_{\mu\nu}^X(t, t_f, t_i; \pvec)  \ =\ \frac{\langle \pi(t_f,\pvec)\,\bar{T}^X_{\mu\nu}(t)\,\pi(t_i,\pvec)\rangle}
  {\langle \pi(t_f,\pvec)\ \pi(t_i,\pvec)}
\end{equation}
which are related to the matrix element
\begin{equation}
  \label{eq:RX}
  R_{\mu\nu}^X(t, t_f, t_i;\pvec) \to
  \frac{1}{2E_\pi}\frac{\langle\pi(\pvec)|\bar{T}^X_{\mu\nu}|\pi(\mathbf{p})\rangle}{1+\exp(-E_\pi(T
    - 2(t_f-t_i)))}
\end{equation}
for $t_f - t$, $t - t_i$ (and thus $t_f - t_i$) large enough such that excited state contributions
have decayed sufficiently. At the same time $T \gsim 2(t_f-t_i)$ should be
maintained, otherwise finite size effects become sizable via excited
states contaminations. $R$ depends on $t_f-t_i$ and $t-t_i$ only, and
in the following we set $t_i=0$.\\
According to \cref{eq:x}, $\avgx$ can be extracted with zero pion
momentum from tensor elements with $\mu=\nu$, whereas for $\mu\neq\nu$ 
nonzero momentum is required. In general, one might expect the signal to be
noisier with nonzero momentum, and this is indeed the case for the
connected-only contribution. However, due to the fact that for
$\mu=\nu$ the signal requires the subtraction of the trace of the
EMT, the quark disconnected and gluon contributions are better
determined from the off-diagonal elements of the EMT, see also
Ref.~\cite{Alexandrou:2020sml}. 
Therefore, we determine the connected-only light contribution to
$\langle x\rangle$ from $\bar{T}_{44}$ at zero pion momentum $\pvec = 0$,
and all the other contributions
from $\bar{T}_{4k}$ with smallest nonzero momentum 
$|\pvec| = 2\pi/L$, averaged
over all six spatial directions.
Further justification for using (off-) diagonal
tensor elements for  (dis-)connected diagrams is given in \seeApp.

For the light-quark connected part both two- and three-point functions
are constructed using
stochastic timeslice sources with spin-color-site components
independently and identically distributed, according to
$\left( \mathbb{Z}_2 + i\, \mathbb{Z}_2 \right)/\sqrt{2}$, 
with random $\mathbb{Z}_2$ noise and eight stochastic samples 
per gauge field configuration. For the quark-disconnected
part of any quark flavour, as well as the gluon operator part,
we employ point-to-all propagators with 200 randomly 
distributed source points per configuration and full spin-color
dilution to estimate the pion two-point function. 
The light-quark loop diagrams with covariant derivative insertion
are determined based on the combination
of low-mode deflation \cite{Gambhir:2016uwp} of the Dirac operator 
with 200 eigenmodes, and hierarchical probing \cite{Stathopoulos:2013aci}
with one stochastic volume source decomposed to coloring distance of
eight lattice sites in each spatial direction. Spin-color dilution is also
employed. The strange quark is treated with the same hierarchical probing setup, but without
deflation. Analogously, for the charm-quark loops we use 12 spin-color diluted volume sources
with coloring distance 4 \cite{Alexandrou:2020sml}.
The gluon field strength tensor in the gluon operator matrix element
is computed with the clover field definition, see
e.g.~\cite{Jansen:1996yt}. We apply ten levels of stout 
smearing~\cite{Morningstar:2003gk} to the gauge links in order to sufficiently
reduce ultraviolet fluctuations, see
Ref.~\cite{Alexandrou:2017oeh,Alexandrou:2020sml}.  

\begin{figure}[t]
  \includegraphics[width=\linewidth]{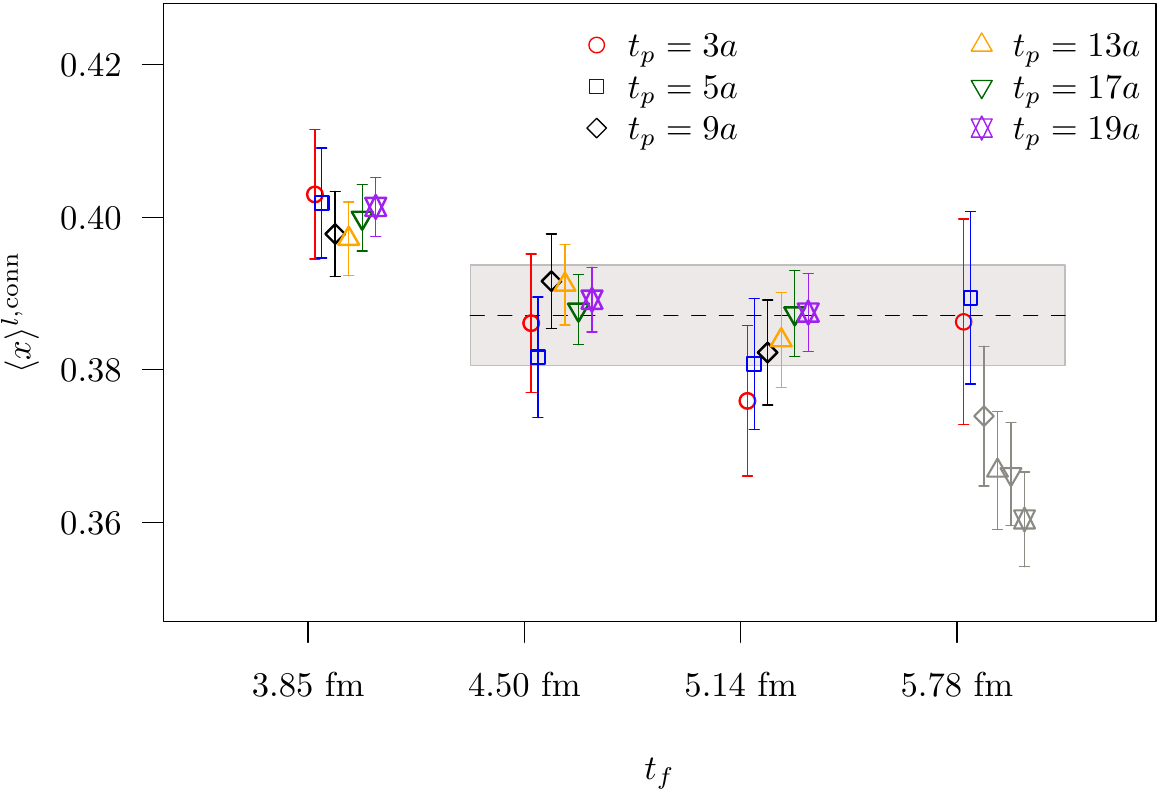}
  \caption{Fit results for the bare light connected $\langle
    x\rangle^{l,\mathrm{conn}}$ as a  function of 
    $t_f$ for different fit range values $t_p$.
    Greyed out points have a $\chi^2/\mathrm{dof}>1.3$. Grey band and
    dashed line represent the result and total error obtained via a
    weighted averaging procedure.}
  \label{fig:ldep}
\end{figure}

Errors are computed using the bootstrap method with fully correlated
fits. Since lattice artefacts are of $\mathcal{O}(a^2)$ with our
discretisation, we expect discretisation errors generically of
the size $a^2\Lambda_\mathrm{QCD}^2$. For the value of the lattice
spacing $a$ used here this amounts to $\sim2.6\%$ with unknown
coefficient. 

\textit{Renormalisation.---}
The quark flavour-nonsinglet combinations renormalise with $Z_{qq}$ as 
\begin{equation}
  \label{eq:nonsinglet}
  \begin{split}
    \langle x\rangle_{u-d}^\mathrm{R} &= Z_{qq}(\langle x \rangle_u - \langle x
    \rangle_d)\,,\\
    \langle x\rangle_{u+d-2s}^\mathrm{R} &= Z_{qq}(\langle x \rangle_u + \langle x
    \rangle_d - 2\langle x\rangle_s)\,,\\
    \langle x\rangle_{u+d+s-3c}^\mathrm{R} &= Z_{qq}(\langle x \rangle_u + \langle x
    \rangle_d + \langle x\rangle_s - 3 \langle x \rangle_c)\,.\\
  \end{split}
\end{equation}
For the pion, $\langle x\rangle_{u-d}^\mathrm{R}=0$ in the isospin
symmetric case, as simulated here.
The quark-singlet and gluon components mix under renormalisation
according to 
\begin{equation}
  \label{eq:mixing}
  \begin{pmatrix}
    \sum_f \langle x\rangle_f^\mathrm{R}\\
    \langle x\rangle_g^\mathrm{R}
  \end{pmatrix}
  =
  \begin{pmatrix}
    Z_{qq}^s & Z_{qg}\\
    Z_{gq} & Z_{gg} \\
  \end{pmatrix}
  \begin{pmatrix}
    \sum_f \langle x\rangle_f\\
    \langle x\rangle_g
  \end{pmatrix}
\end{equation}
with $Z_{qq}^s$ the quark-singlet renormalisation constant.
Defining $\delta Z_{qq} = Z_{qq}^s - Z_{qq}$,  one can solve 
the set of Eqs.~(\ref{eq:mixing}) for each single flavour and gluon
component: 
\begin{equation}
  \label{eq:separate}
  \begin{split}
    \langle x\rangle_f^\mathrm{R} &= Z_{qq}\langle x\rangle_f +
    \frac{\delta Z_{qq}}{N_f}\sum_{f'} \langle x\rangle_{f'} +
    \frac{Z_{qg}}{N_f}\langle x \rangle_g\,,\\
    \langle x\rangle_g^\mathrm{R} &= Z_{gg}\langle x\rangle_g + Z_{gq}\sum_{f'} \langle x\rangle_{f'}\,.\\
  \end{split}
\end{equation}
Due to lattice artefacts, renormalisation factors are different for
$\bar{T}_{\mu\nu}$ with $\mu=\nu$ and $\mu\neq\nu$.  
The diagonal elements of the renormalisation matrix have been
determined nonperturbatively and the off-diagonal elements
perturbatively in Ref.~\cite{Alexandrou:2020sml}, see also \seeApp. 
Since these mixing coefficients have been determined using one-loop
perturbation theory, we do not have an error estimate available. In
order to account for the uncertainty, we perform the computation once
including the mixing, and once excluding it, and take the spread as
error estimate.

\textit{Results.---}
We compute $R^X(t)$ in \cref{eq:Rdef} for various values of
$t_f$. Solving \cref{eq:x} for $\avgx$ and inserting \cref{eq:RX},
we then extract $\avgx^X(t)$, where $X$ stands for $l,\mathrm{conn}$
(with $l\equiv u+d$), $l,\mathrm{disc}$, $s$, $c$, and $g$.
For large enough $t_f$ we expect $\avgx(t)$ to show a plateau for
$t-t_f/2$ around $0$.
Thus, we fit a constant symmetrically around $t-t_f/2=0$ with
fit range denoted as $t_p$ to our bare data for $\avgx(t)$ (for plots
of this bare data see \seeApp). In
\cref{fig:ldep} we show the result of 
such constant fits to the light connected contribution as a function
of the source sink separation 
$t_f$ for different values of $t_p$. Between
$t_f=4.5\ \mathrm{fm}$ and $t_f=5.14\ \mathrm{fm}$ we see
agreement for all values of $t_p$. For $t_f=5.78\ \mathrm{fm}$ the
results for the smallest three $t_p$ values still agree with the
previous ones. However, for the larger $t_p$ values we start to see
finite size effects due to $T/2<t_f$, also visible in the bad
$\chi^2/\mathrm{dof}$ values. 

\begin{figure}[t]
  \includegraphics[width=\linewidth]{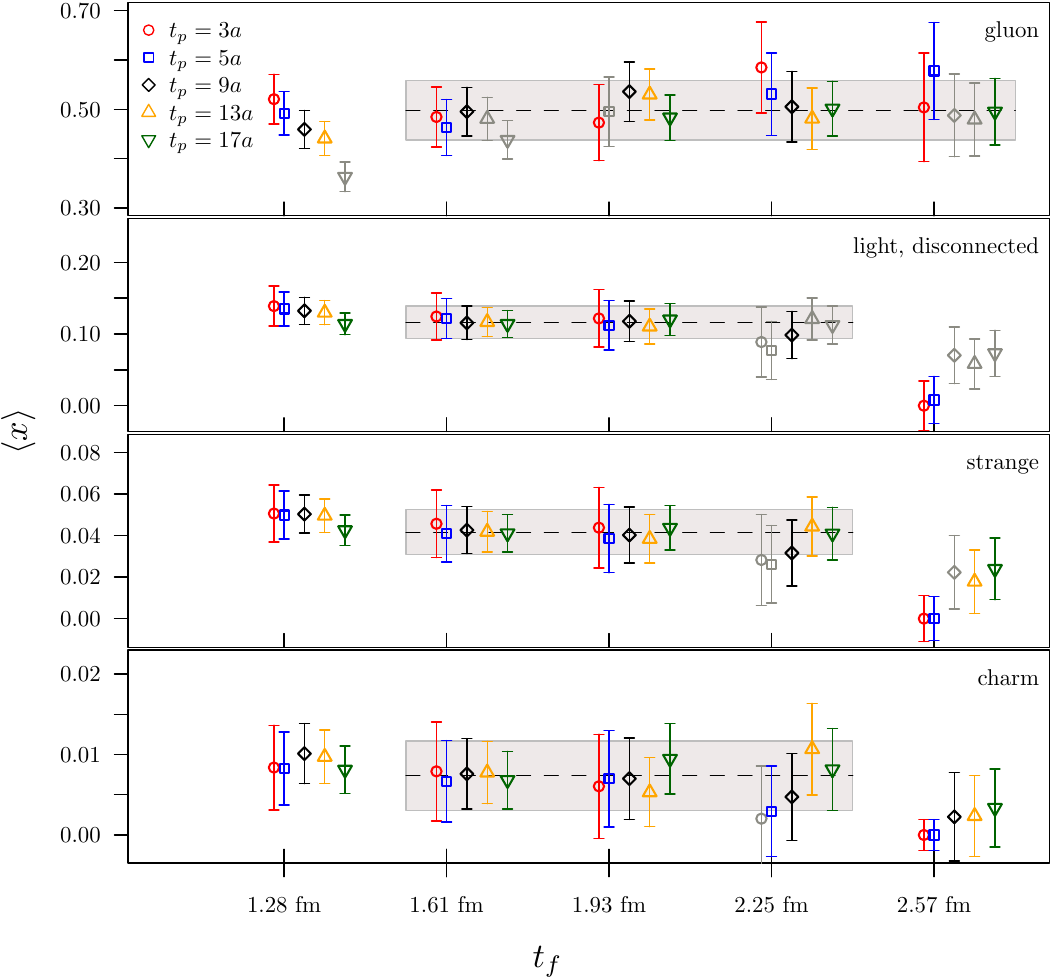}
  \caption{Fit results for the bare quark disconnected and the
    gluon $\langle x \rangle$ as a  function of 
    $t_f$ for different fit range values $t_p$.
    Greyed out points have a $\chi^2/\mathrm{dof}>1.3$. Grey band and
    dashed line represent the result and total error obtained via a
    weighted averaging procedure.}
  \label{fig:depcomb}
\end{figure}

In \cref{fig:depcomb} we again show the fit results as a function of
$t_f$ for different $t_p$ values, but here for the light
disconnected and the strange, charm and gluon contributions. For the
quark disconnected contributions we loose the signal for
$t_f>2.25\ \mathrm{fm}$. However, for all three cases we observe
agreement between all results for $1.61\ \mathrm{fm} \leq t_f\leq
2.25\ \mathrm{fm}$, confirming ground state dominance.
Thus, we are confident that
the final result can be determined in this region of $t_f$ values.

We arrive at the final result by assigning a weight
$w\ =\ \exp\left(-\frac{1}{2}\left[\chi^2 - 2\,\mathrm{dof}\right]\right)$
to every fit with given $\chi^2$ value and degrees of freedom
($\mathrm{dof}$). Then we take the weighted average (see
also~\cite{Borsanyi:2020mff}) 
over all constant fits in the aforementioned regions of $t_f$
values. The \emph{combined statistical and systematic} error is
computed by repeating this procedure on all bootstrap samples with
weights corresponding to the fits on the samples.
Alternatively, we have also performed fits which take explicitly into
account excited state contaminations again for various $t_f$ and
$t_p$ values leading to consistent results, but with a
different  distribution of statistical and systematic errors.

\begin{figure}
  \includegraphics[width=\linewidth]{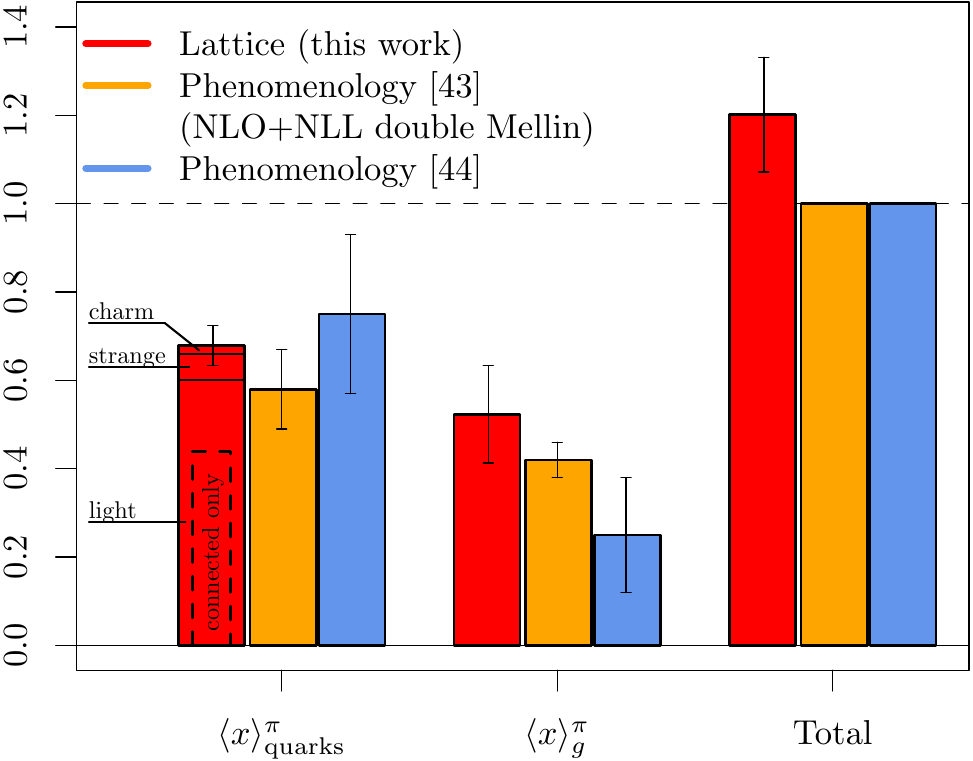}
  \caption{Comparison to Ref.~\cite{Barry:2021osv} and
    Ref.~\cite{Novikov:2020snp} at 
    $2\ \mathrm{GeV}$ in the $\overline{\mathrm{MS}}$ scheme. In the
    two phenomenological computations the momentum sum rule is imposed.}
  \label{fig:comp}
\end{figure}

Using the so extracted bare values for $\langle x\rangle^X$ (see
\seeApp), we are  now in the position to compute the renormalised 
flavour nonsinglet and singlet contributions to $\langle x\rangle$. We obtain
for the nonsinglet ones from \cref{eq:nonsinglet}
\begin{equation}
    \langle x\rangle^\mathrm{R}_{u+d-2s} = 0.48(1)\,,\
    \langle x\rangle^\mathrm{R}_{u+d+s-3c} = 0.60(3)\,,
\end{equation}
where we recall that $\langle x\rangle_{u-d}=0$ due to isospin symmetry in
the light-quark sector. For the singlet contributions we find, using \cref{eq:mixing,eq:separate},
\begin{equation}
    \sum_f\langle x \rangle^\mathrm{R}_f = 0.68(5)(_{-3})\,,\
    \langle x \rangle^\mathrm{R}_g = 0.52(11)(^{+2})\,.
\end{equation}
The first error represents  the combined statistical and fit range
uncertatinty, the second error comes from the mixing under renormalisation.
The sum of all contributions amounts to $\langle
x\rangle^\mathrm{R}_\mathrm{total} = 1.20(13)(_{-3})$, compatible
with the expected value of $1$ within two sigma.
This is
an important result because, in contrast to phenomenological analyses where
the saturation of the momentum sum rule is imposed, in our work
such saturation is a result of the computation.
In \cref{fig:comp} we compare to the recent phenomenological
results\footnote{We thank the authors of
  Ref.~\cite{Barry:2021osv} for
  communicating their results at $2\ \mathrm{GeV}$ in the
  $\overline{\mathrm{MS}}$ scheme.}
from Ref.~\cite{Barry:2021osv} and Ref.~\cite{Novikov:2020snp}. 
Our work agrees within errors with these state-of-art phenomenological 
results. 

In \cref{tab:results} we compile all the contributions again and
compare to the literature. The only other lattice result was presented
in Ref.~\cite{Loffler:2021afv}, where  
$N_f=2+1$ flavour QCD was used and results have been extrapolated to
the continuum and the physical point. However, the gluon contribution
has not been computed and, thus, the mixing could not be taken into
account. Therefore, the comparison of the quark singlet contributions
is of limited meaningfulness. The nonsinglet contribution $\langle
x\rangle^\mathrm{R}_{u+d-2s}$ is better suited for a comparison. 
While currently there is a discrepancy between the results, we notice that 
a full comparison should be attempted only after the inclusion of 
all unaccounted systematics.

Finally, the results from the momentum sum rule decomposition can 
be used to determine how quarks and gluons contribute to the pion 
mass. In principle, mass can be decomposed in QCD in different 
ways~\cite{Ji:1994av,Ji:1995sv,Lorce:2017xzd,Hatta:2018sqd,Tanaka:2018nae,Metz:2020vxd}.
Choosing the sum rule of Ref.~\cite{Metz:2020vxd},  $M_{\pi,q} = (3
M_\pi/4)\langle x\rangle^\mathrm{R}_\mathrm{quarks}$ and
$M_{\pi,g} = (3 M_\pi/4)\langle x\rangle^\mathrm{R}_{g}$, which amounts
to $70(5)\ \mathrm{MeV}$ and $55(12)\ \mathrm{MeV}$ at
$2\ \mathrm{GeV}$ in the $\overline{\mathrm{MS}}$ scheme,
respectively. Note that the gluon contribution is the same in Ji's
original mass decomposition~\cite{Ji:1994av,Ji:1995sv}. The remaining
contribution is split among a trace anomaly term and a term
proportional to the quark mass.

\begin{table}[th]
  \centering
  \begin{tabular*}{.49\textwidth}{@{\extracolsep{\fill}}lllll}
    \toprule\hline
    & this work & \cite{Loffler:2021afv} & \cite{Barry:2021osv} & \cite{Novikov:2020snp}\\
    \midrule\hline
    \(\langle x\rangle_l^\mathrm{R}\) & $0.601(28)(_{-21})$ & -- & -- & --\\
    \(\langle x\rangle_s^\mathrm{R}\) & $0.059(13)(_{-10})$ & -- & -- & --\\
    \(\langle x\rangle_c^\mathrm{R}\) & $0.019(05)(_{-10})$ & -- & -- & --\\
    \(\langle x\rangle_g^\mathrm{R}\) & $0.52(11)(^{+02})$ & -- & 0.42(4) & 0.25(13)\\
    \(\sum_f\langle x\rangle_f^\mathrm{R}\) & $0.68(05)(_{-03})$ & 0.220(207) & 0.58(9) & 0.75(18)\\
    \(\langle x \rangle^\mathrm{R}_{u+d-2s}\) & $0.48(01)$ & 0.344(28) & -- & --\\
    \(\langle x \rangle^\mathrm{R}_{u+d+s-3c}\) & $0.60(03)$ & -- & -- & --\\
    \bottomrule\hline
  \end{tabular*}
  \caption{Compilation of results and comparison to literature. All
    values are at $2\ \mathrm{GeV}$ in the $\overline{\mathrm{MS}}$
    scheme.}
  \label{tab:results}
\end{table}

\textit{Summary and Outlook.---}
In this letter we have presented results for the complete flavour
decomposition of the average momentum of quarks and gluons in the
pion for the first time. The computation in $N_f=2+1+1$ lattice QCD
are performed directly with physical values of the quark mass
parameters making an extrapolation to the physical point superfluous
and, thus, avoiding any systematic uncertainty from such an
extrapolation. However, we work at a single value of the lattice
spacing only, which does not allow us to take the continuum
limit. Therefore, we have to expect lattice artefacts of
$\mathcal{O}(a^2)$ which we cannot account for rigorously.
The renormalisation constants have been computed nonperturbatively,
while the mixing coefficients were computed in perturbation theory.

We find the momentum sum rule to be fulfilled within two sigma
errors, see \cref{fig:comp}. When comparing to phenomenological
determinations from 
Refs.~\cite{Novikov:2020snp,Barry:2021osv} we find reasonable
agreement within relatively large uncertainties. Comparing to the only
other lattice QCD computation~\cite{Loffler:2021afv} including quark
disconnected contributions, but not the mixing and the gluon
contribution, we observe a deviation well ouside uncertainties. 

Future plans include determining $\langle x\rangle$ in the pion
for two more lattice spacing values directly at the physical
point. Preliminary results for the flavour non-singlet components at a
finer lattice spacing show agreement within errors.
Moreover, work is in progress to determine the mixing
coefficients nonperturbatively. This work opens the possibility to
combine lattice QCD results and experimental data in a global
phenomenological analysis.

\begin{acknowledgments}
  \textit{Acknowledgements.---}
  The authors thank M.~Constantinou for very helpful discussions and
  gratefully acknowledge the Gauss Centre for
  Supercomputing e.V. (www.gauss-centre.eu) for funding this project
  by providing computing time on the GCS Supercomputer
  JUQUEEN~\cite{juqueen} and the John von Neumann Institute for
  Computing (NIC) for computing time provided on the supercomputers
  JURECA~\cite{jureca} and JUWELS~\cite{juwels} at Jülich
  Supercomputing Centre (JSC).  We further acknowledge computing time
  granted on Piz Daint at Centro Svizzero di Calcolo Scientifico
  (CSCS) via the project with ids s849, s702 and s954.
This work is supported in part by the Deutsche Forschungsgemeinschaft (DFG,
  German Research Foundation) and the 
  NSFC through the funds provided to the Sino-German
  Collaborative Research Center CRC 110 “Symmetries
  and the Emergence of Structure in QCD” (DFG Project-ID 196253076 -
  TRR 110, NSFC Grant No.~12070131001), the Swiss National Science
  Foundation (SNSF) through grant No.~200021\_175761 and the Marie
  Skoldowska-Curie European Joint Doctorate STIMULATE funden by the
  European Union's Horizon 2020 research \& innovation program under
  Grant agreement No. 765048. FP acknowledges financial support from 
  the Cyprus Research and Innovation Foundation under project ''NextQCD``,
  contract no. EXCELLENCE/0918/0129.
The open source software packages tmLQCD~\cite{Jansen:2009xp,Abdel-Rehim:2013wba,Deuzeman:2013xaa}, 
  Lemon~\cite{Deuzeman:2011wz}, 
  QUDA~\cite{Clark:2009wm,Babich:2011np,Clark:2016rdz} and
  R~\cite{R:2019,hadron:2020} have  
  been used.
\end{acknowledgments}

\section{Appendix}

\begin{figure}[t]
  \includegraphics[width=\linewidth]{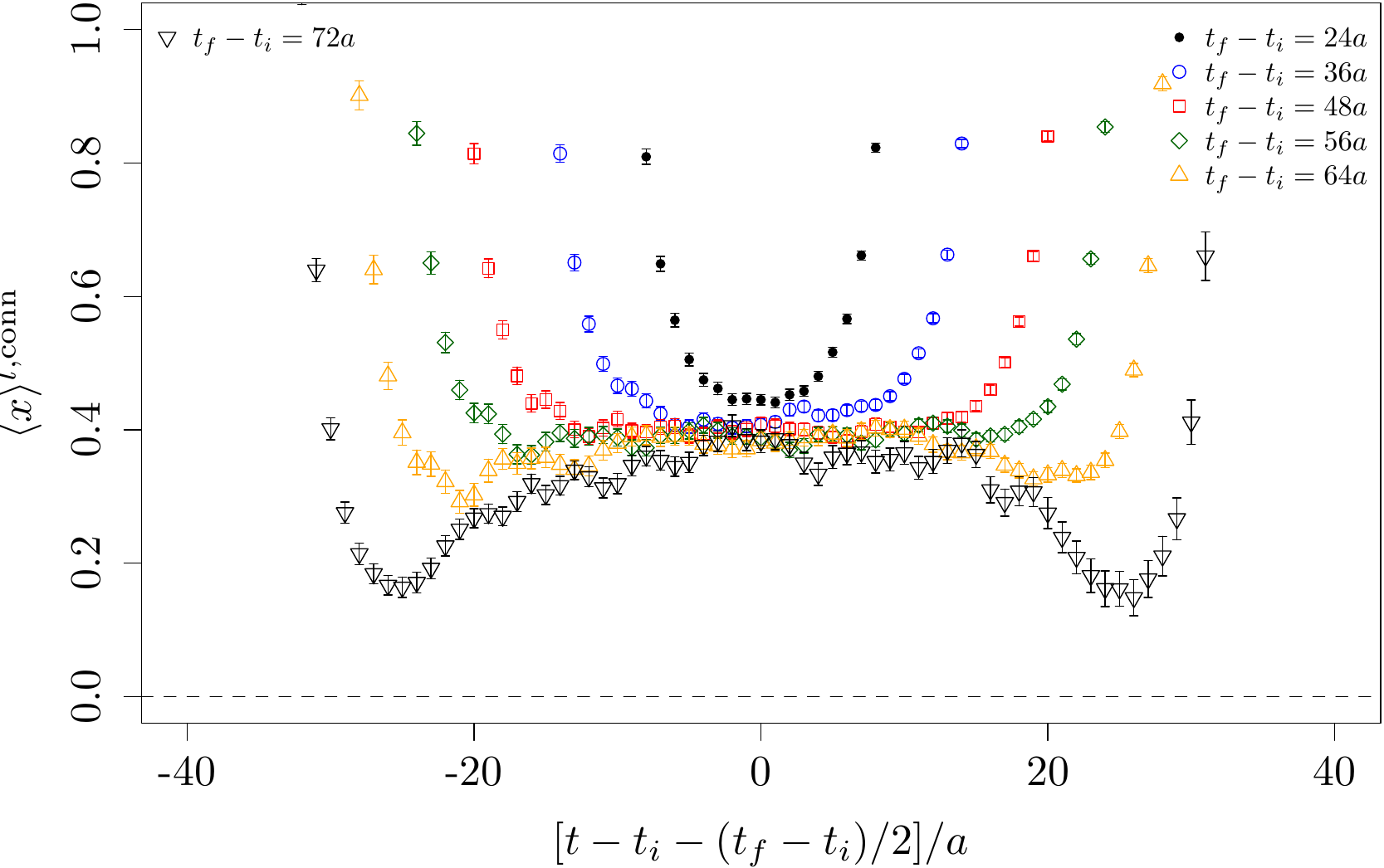}
  \caption{Connected
    contribution to the bare $\langle x\rangle^l$ is shown for
    different values of $t_f-t_i$.}
  \label{fig:connratios}
\end{figure}

In this section we provide additional details on the analysis and
additional figures. In \cref{fig:connratios} we show
$\avgx^{l,\mathrm{conn}}(t-t_i)$ for different source sink separations $t_f-t_i$. It
is clearly observable that the plateau around $t-t_i-(t_f-t_i)/2=0$
becomes longer as $t_f-t_i$ increases. However, once
$t_f-t_i>T/2$, deviations become larger again.

\begin{figure}
  \includegraphics[width=\linewidth]{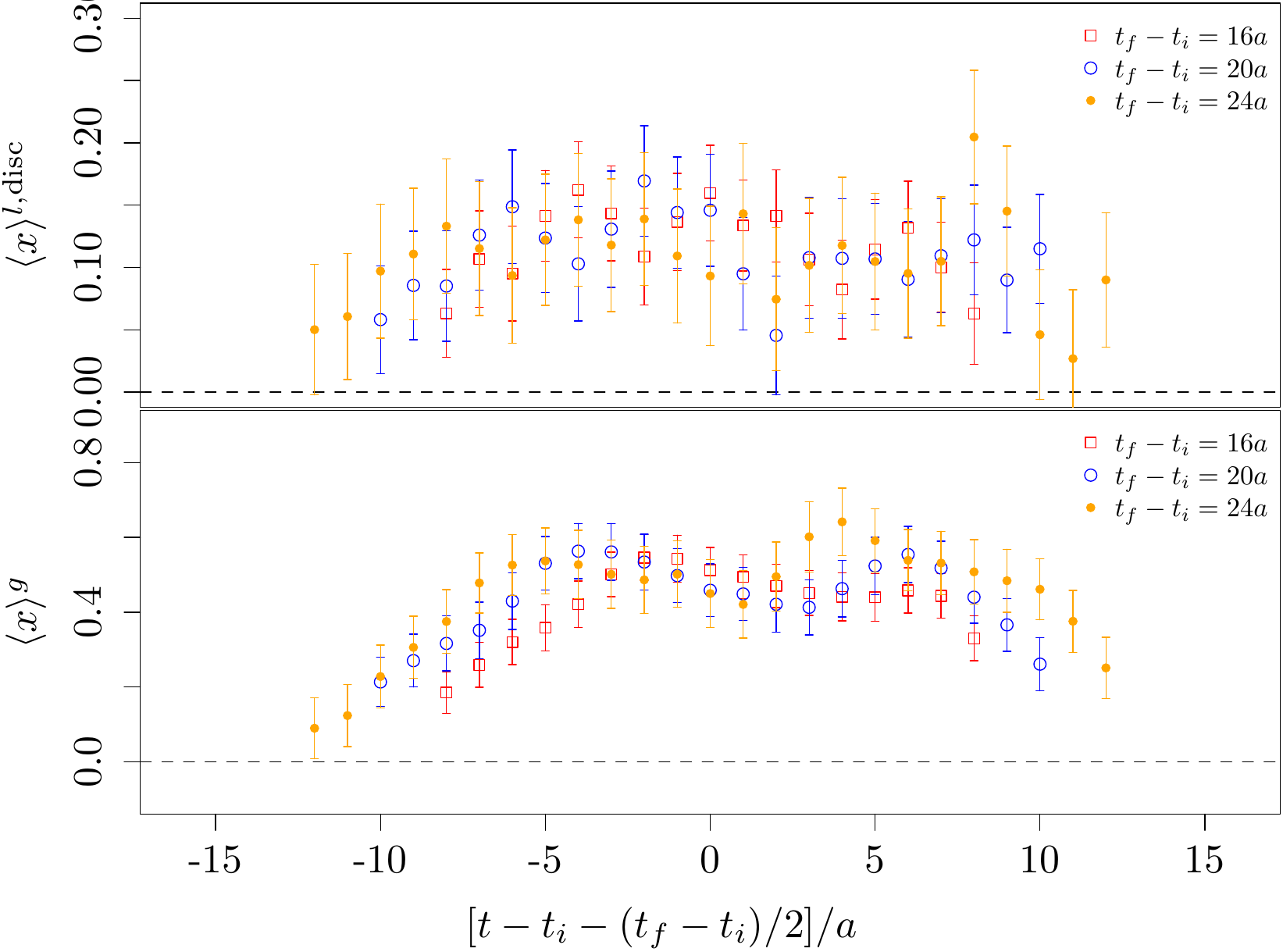}
  \caption{Bare $\langle x\rangle^X$ as a function of $t-t_i$ for $X=l$
    (disconnected only) and $X=g$. Different values of $t_f-t_i$ are
    shown and we only show $t\in[0,t_f-t_i]$.}
  \label{fig:discratios}
\end{figure}

For the disconnected contributions we show as examples $\avgx^g(t-t_i)$
and $\avgx^{l,\mathrm{disc}}$ in \cref{fig:discratios}. It can be seen
that the plateau region around $t-t_i-(t_f-t_i)/2=0$ is rather independent
of the shown $t_f-t_i$ values, apart from increasing statistical
uncertainties with increasing $t_f-t_i$ values. For $t_f-t_i=32$ (not
shown) errors are so large, that all points are compatible with zero
for the fermionic disconnected contributions. 

In \cref{tab:bareresults} we give the bare values for all the
contributions to $\avgx$.

\begin{table}[th]
  \centering
  \begin{tabular*}{.49\textwidth}{@{\extracolsep{\fill}}lll}
    \toprule\hline
    Contribution & Operator &\\
    \midrule\hline
    \(\langle x\rangle_l^\mathrm{conn}\) & \(\bar{T}_{44}\) & \(0.387(7)\) \\
    \(\langle x\rangle_l^\mathrm{disc}\) & \(\bar{T}_{4k}\) & \(0.116(23)\) \\
    \(\langle x\rangle_s\)              & \(\bar{T}_{4k}\) & \(0.0416(108)\) \\
    \(\langle x\rangle_c\)              & \(\bar{T}_{4k}\) & \(0.0074(43)\) \\
    \(\langle x\rangle_g\)              & \(\bar{T}_{4k}\) & \(0.498(61)\) \\
  \bottomrule\hline
  \end{tabular*}
  \caption{Bare values for the different contributions to $\avgx$.}
  \label{tab:bareresults}
\end{table}

\subsection{Renormalisation}
\label{subsec:discon}

For convenience we reproduce here the renormalisation constants from
Ref.~\cite{Alexandrou:2020sml}. The nonsinglet ones read
\[
Z_{qq}^{\mu=\nu}\ =\ 1.151(1)(4)\,,\quad
Z_{qq}^{\mu\neq\nu}\ =\ 1.160(1)(3)\,,
\]
and the singlet ones
\[
Z_{qq}^{s,\mu=\nu}=1.161(18)(16)\,,\quad
Z_{qq}^{s,\mu\neq\nu}=1.163(11)(5)\,.
\]
For the gluon
\[
Z_{gg}^{\mu\neq\nu}\ =\ 1.08(17)(3)\, ,
\]
all at $2\ \mathrm{GeV}$ in the $\overline{\mathrm{MS}}$ scheme.
The mixing coefficients $Z_{gq}$ and $Z_{qg}$ have been determined
perturbatively in Ref.~\cite{Alexandrou:2020sml}. Their values read
\begin{equation}
  \label{eq:Zmix}
  \begin{split}
    Z_{gq}^{\mu=\nu}\ &=\ 0.232\,,\quad Z_{gq}^{\mu\neq\nu}\ =\ 0.083\,,\\
    Z_{qg}^{\mu=\nu}\ &=\ -0.027\,.\\
  \end{split}
\end{equation}
Unfortunately, $Z_{qg}$ is not available for ${\mu\neq\nu}$. However,
since $Z_{gq}^{\mu\neq\nu}\ll Z_{gq}^{\mu=\nu}$ and $Z_{qg}^{\mu=\nu}$
itself is small compared to the other two, we assume here that
$Z_{qg}^{\mu=\nu}=Z_{qg}^{\mu\neq\nu}$, also due to the fact that the
divergent part in the two is identical.

\newcommand{\chibar}{\bar{\chi}}
\newcommand{\Tbar}{\bar{T}}
Another question which needs to be discussed is the justification of
treating connected and disconnected light contributions separately and
even using different operators. It is based on the
mixed action analysis performed by the authors of
Ref.~\cite{Frezzotti:2004wz}. 

Consider the bare quark contribution to $\avgx$ for a twisted mass quark
flavor $\chi$ not present in the simulated action with  
twisted quark mass $\mu_\chi$. The insertion $\Tbar^{\chi}$ leads to 
a purely disconnected contribution. For such an insertion, at nonzero
lattice spacing the corresponding $\avgx^{\chi}(\mu_\chi)$ is well
defined from spectral decomposition. Lattice rotational symmetry instead
guarantees \emph{invariance} of $\avgx^{\chi}(\mu_\chi)$ w.r.t. the
chosen tensor component \emph{up to lattice artifacts}.

We use this \emph{nonunitary} setup for the strange and charm quark
contribution with any value of $\mu_s,\,\mu_c$. Note that lattice
artefacts are then still of $\mathcal{O}(a^2)$~\cite{Frezzotti:2004wz}. 
In the regularised theory the so obtained $\avgx^{\chi}(\mu_\chi)$
is a \emph{smooth function} of the twisted valence quark mass
$\mu_\chi$ and we can have a well-defined limit to the unitary case 
\begin{align}
  \lim\limits_{\mu_\chi \to \mu_l}\,\avgx^{\chi}_{\mu_\chi} &= \avgx^{l,\mathrm{disc}} \,,
  \label{eq:discon-1}
\end{align}
where results from different choices of tensor components only differ
by lattice artefacts. Thus in the linear decomposition of $\avgx =
\avgx^{l,\mathrm{conn}} + \avgx^{l,\mathrm{disc}}$ into a connected
and disconnected contribution, the disconnected contribution
$\avgx^{l,\mathrm{disc}}$  is fixed up to lattice artefacts.
Also the complete $\avgx^{l,\mathrm{conn}} + \avgx^{l,\mathrm{disc}}$
is invariant under change of tensor components by the remnant Lorentz
symmetry.

In conclusion the employed decomposition into $\avgx^{l,\mathrm{conn}}
+ \avgx^{l,\mathrm{disc}}$ in the regularised theory is unique up to
lattice artefacts. Moreover, the renormalisation pattern for diagonal
and off-diagonal tensor elements is identical again up to lattice
artefacts, which vanish in the continuum limit.

Note that the spectral decomposition of the relevant three-point
function leads to a constant plus excited states. This is why there
cannot be any delicate cancellation of exponential terms like in the
case for instance of the $\eta,\eta^\prime$ correlation functions,
where connected and disconnected contribution must not be analysed
separately. 

\subsection{Quark mass tuning}

The unitary light, strange and charm quark masses for the simulated lattice action
are tuned such that the physical values of the pion mass, the ratio of charm and strange quark masses and the ratio of the $D_s$ meson and
its decay constant are reproduced as detailed in Ref.~\cite{Alexandrou:2018egz}.

Within our mixed action setup, the bare mass of strange and charm quark, which enter
the quark energy momentum tensor in our calculation, are tuned by matching  the $\Omega^-$ and the $\Lambda_c$ baryon mass
to their physical values by the procedure in Ref.~\cite{Alexandrou:2021gqw}.

The numerical values of the bare strange and charm quark parameters read
\begin{align}
  a\, \mu_s = 0.0186 \,,\quad a\, \mu_c = 0.2490 \,.
  \label{eq:OS-masses}
\end{align}

\end{document}